\documentclass[preprint]{revtex4}%
\usepackage{graphicx}
\usepackage{dcolumn}
\usepackage{bm}
\usepackage{amsmath}
\usepackage{amsfonts}
\usepackage{amssymb}%
\begin{document}
\title{Screened Coulomb interaction in the maximally localized Wannier basis}
\author{Takashi Miyake and F. Aryasetiawan}
\affiliation{Research Institute for Computational Sciences, AIST, Tsukuba 305-8568, Japan}
\affiliation{Japan Science and Technology Agency, CREST}
\date{\today}

\begin{abstract}
We discuss a maximally localized Wannier function approach for constructing
lattice models from first-principles electronic structure calculations, where
the effective Coulomb interactions are calculated in the constrained
random-phase-approximation. The method is applied to the 3d transition metals
and a perovskite (SrVO$_{3}$). We also optimize the Wannier functions by
unitary transformation so that \emph{U} is maximized. Such Wannier functions
unexpectedly turned out to be very close to the maximally localized ones.

\end{abstract}

\maketitle

\section{Introduction}

In a class of materials often referred to as correlated materials, the
electronic structure is characterized by a set of partially filled narrow
bands across the Fermi level. Seen from the atomic site, one has a set of
partially filled shell of localized orbitals typically of 3d or 4f character.
Many of the electronic properties of the material are determined by the
correlations among the localized electrons living in the partially filled band
or shell. It is therefore physically well motivated to map the original
complicated many-electron problem to a model consisting of the localized
orbitals and a few additional orbitals. By eliminating the high-energy states
("downfolding") the long-range bare Coulomb interaction is screened to a
short-range interaction at low energy. Since the screened interaction is short
range, only on-site interaction or the Hubbard \emph{U} is often taken into
consideration in the model. This is the physical idea behind the well-known
Hubbard model or Andersen impurity model. One would then wish to have a set of
well localized orbitals or Wannier orbitals that span the same Hilbert space
as that of the states that form the narrow bands. In this way the Hubbard
\emph{U} will have small off-site matrix elements, which may be neglected.
Practical procedures to construct the models starting from first-principles
calculations have been a subject of interest for a long time
\cite{gunnarsson89,gunnarsson90,aryasetiawan04,solovyev06}.

In this work, we focus on Wannier orbitals using the method developed by
Souza, Marzari and Vanderbilt \cite{marzari97,souza01} based on the
minimization of the quadratic extent of the orbitals. An alternative, equally
promising approach is to use the Wannier orbitals of Andersen
\cite{andersen00}. While the former is a "post-processing" method, i.e., the
Wannier orbitals are constructed after generating the Bloch wave functions,
the latter may be termed "pre-processing" method because the Wannier orbitals
are constructed before diagonalization of the Hamiltonian that yields the band
structure \cite{marzari97}. In this sense, the latter scheme may be more
advantageous than the former. On the other hand, the former is more general
because it does not depend on any particular band-structure method. Comparison
between the two Wannier functions for some selected materials can be found in
Ref.\cite{lechermann06}.

Apart from the use of Wannier orbitals in constructing lattice models, there
are many other applications. In particular, a close connection with Berry's
phase \cite{kingsmith93,marzari97} has stimulated intensive works recently
\cite{giustino06,wang06,yates07,miyake07}. Related work can be found not only
in condensed matter physics but also in chemistry, where the concept of
localized molecular orbitals is very useful for understanding chemical bonding
as well as for visualization. The idea of constructing localized molecular
orbitals goes back to the early sixties by the maximization of the Coulomb
energy of the molecular orbitals \cite{edmiston63} or the minimization of the
quadratic extent of the molecular orbitals \cite{boys66}. The general problem
of transforming a set of Bloch states to a set of well localized orbitals is
therefore one of important methodological problems in condensed matter physics
and chemistry.

Another important issue in the downfolding procedure is how to determine
effective interaction parameters. A widely used method is constrained LDA
(cLDA) \cite{gunnarsson89,gunnarsson90,anisimov91} and a recently proposed
scheme based on the maximally localized Wannier function \cite{nakamura06} may
be useful for applications to complicated structures. On the other hand, cLDA
is known to yield unreasonably large values of $U$ in some cases (e.g. late
transition metals). This arises from technical difficulty in including part of
the self-screening processes between localized electrons leading in some cases
to a larger value of \emph{U} \cite{aryasetiawan06}. Another method for
estimating effective interaction is the random phase approximation (RPA). We
can find early trials along this line in Ref.\cite{springer98,kotani00}. Later
on the constrained RPA (cRPA) scheme was invented \cite{aryasetiawan04}. The
cRPA method has several advantages over currently available methods. It allows
for a precise elimination of screening channels, which are to be included in a
model Hamiltonian, without modifying the one-particle dispersion of the model.
In addition, the effective screened interaction as a function of \textbf{r}
and \textbf{r$^{\prime}$} can be calculated independent of the basis functions. We will
use this method in the present work. We can also find other proposals in
literature such as a hybrid method between cLDA and cRPA \cite{solovyev05} and
linear response approach \cite{cococcioni05}.

Our long-term goal is to construct a first-principles scheme for calculating
the electronic structure of correlated materials. As is well known, the local
density approximation (LDA) \cite{kohn65} in density functional theory (DFT)
\cite{hohenberg64} often has difficulties when applied to such systems.
Attempts to improve the LDA have resulted, among others, in the LDA+U
\cite{anisimov91,anisimov97a,anisimov01}, the LDA+DMFT (Dynamical Mean-Field
Theory \cite{georges92,georges96}) \cite{anisimov97b} and more recently in the
newly developed GW+DMFT method \cite{sun02,biermann03}. In
Ref.\cite{biermann03} it is shown how the Hubbard \emph{U} for real materials
can be determined self-consistently within the scheme. In these methods it is
crucial to have well localized orbitals representing the Hilbert space of the
partially filled correlated bands since the screened Coulomb interaction
\emph{U} is usually assumed to be purely on-site. This is especially the case
in the DMFT method, where the lattice problem is mapped to an impurity problem
\cite{paul06}.

The purpose of the present work is to demonstrate the usefulness of the
Wannier orbitals and the feasibility of performing many-body calculations with
a unified Wannier basis, independent of the starting band structure. To this
end we have calculated the Hubbard \emph{U} within the cRPA scheme using the
maximally localized Wannier basis and compared the results with independent
calculations \cite{aryasetiawan04,aryasetiawan06} in the linear muffin-tin
orbital (LMTO) basis \cite{andersen75}. The reasonably good agreement between
the two sets of results gives us confidence as to the usefulness of the scheme.

As mentioned earlier, it is highly desirable to construct a set of Wannier
orbitals that minimize the off-site Coulomb interaction or equivalently
maximizes the on-site \emph{U}. We follow the method of Edmiston and
Ruedenberg \cite{edmiston63}, which was proposed for molecules, and developed
a practical procedure to maximize the \emph{U} parameter for periodic crystals
through unitary transformation in real space. Application to transition metals
shows that the effect of maximization is tiny if we start the optimization
from the maximally localized Wannier functions.

\section{Method}

The Wannier function with band index $n$ at cell $\mathbf{R}$ is defined by%

\begin{equation}
| \varphi_{n\mathbf{R}} \rangle= \frac{V}{(2\pi)^{3}} \int e^{-i\mathbf{k}%
\cdot\mathbf{R}} | \psi_{n\mathbf{k}}^{\mathrm{(w)}} \rangle d^{3}k\;,
\label{eq:wannier}%
\end{equation}
where $| \psi_{n\mathbf{k}}^{\mathrm{(w)}} \rangle$ is the associated Bloch
function which can be expanded as a linear combination of the eigenfunctions
of a mean-field Hamiltonian as%

\begin{equation}
| \psi_{n\mathbf{k}}^{\mathrm{(w)}} \rangle= \sum_{m} {\mathcal{U}}%
_{mn}(\mathbf{k}) | \psi_{m\mathbf{k}} \rangle\;. \label{eq:umn}%
\end{equation}
In practical implementations, Kohn-Sham wavefunctions may be used for $|
\psi_{m\mathbf{k}} \rangle$. In the maximally localized Wannier function
scheme \cite{marzari97,souza01}, the coefficients ${\mathcal{U}}%
_{mn}(\mathbf{k})$'s are determined such that the quadratic extent of wavefunctions%

\begin{equation}
\Omega= \sum_{n} ( \langle\varphi_{n \mathbf{0}} | r^{2} | \varphi_{n
\mathbf{0}} \rangle- | \langle\varphi_{n \mathbf{0}} | \mathbf{r} | \varphi_{n
\mathbf{0}} \rangle|^{2} ) \;,
\end{equation}
is minimized. For this purpose, we introduce an energy window and optimize
$\mathcal{U}_{mn}(\mathbf{k})$ with limiting $m$ to the states inside the
window. The parameters for this window ("window 1") are listed in Table
\ref{tab:window}. The Wannier function is more localized as the energy window
is larger, since optimization is done in wider Hilbert space. We found,
however, that the (screened) Coulomb interaction is not sensitive to the
choice of the energy window unless the window is too wide.

\begin{table}[ptb]
\caption{Energy window 1 that limits states to be included in constructing
Wannier functions, and window 2 to specify localized orbitals $\{ \psi_{d} \}$
for cRPA. The "2nd band" for the window 2 means the second lowest one in the
bands with strong 4$s$ and 3$d$ character. Energies are measured from the
Fermi level. }%
\label{tab:window}
\begin{ruledtabular}
\begin{tabular}{l c c}
& Window 1 & Window 2 \\
\hline
Sc & [-3.0 eV, 5.0 eV] & [2nd band, 4.05 eV] \\
Ti & [-4.0 eV, 5.0 eV] & [2nd band, 3.85 eV] \\
V & [-4.0 eV, 5.0 eV] & [2nd band, 4.15 eV] \\
Cr & [-5.0 eV, 4.0 eV] & [2nd band, 2.85 eV] \\
Mn & [-5.0 eV, 4.0 eV] & [2nd band, 1.50 eV] \\
Fe & [-5.0 eV, 4.0 eV] & [2nd band, 1.20 eV] \\
Co & [-5.0 eV, 3.0 eV] & [2nd band, 0.55 eV] \\
Ni & [-7.0 eV, 3.0 eV] & [2nd band, 0.25 eV] \\
SrVO$_3$ & [-10.0 eV, 5.0 eV] & 3 $t_{2g}$ states \\
\end{tabular}
\end{ruledtabular}
\end{table}

The idea of cRPA is to define an effective interaction $W_{r}$ by excluding
screening processes that are included in an effective low-energy model with
$W_{r}$ as the effective interaction (Hubbard \emph{U}). To this end we divide
the Hilbert space into two parts: localized states that form the projected
bands $\{\psi_{d}\}$ and the rest $\{\psi_{r}\}$. The polarizability $P$ is
then divided into two as $P=P_{d}+P_{r}$, where $P_{d}$ includes transitions
between $\{\psi_{d}\}$ only and $P_{r}$ is the rest of the polarization. The
effective interaction in the reduced space $W_{r}$ is defined so that
$W_{r}[1-P_{d}W_{r}]^{-1}$ yields the RPA fully screened interaction
$W=v[1-Pv]^{-1}$, where $v$ is the bare Coulomb interaction. It can be shown
that such $W_{r}$ is given by%

\begin{equation}
W_{r}(\omega)=v[1-P_{r}(\omega)v]^{-1}\;,
\end{equation}
where spatial coordinates are omitted for simplicity \cite{aryasetiawan04}.
This $W_{r}$, after multiplying some localized functions and integrating over
space, can be interpreted as the frequency-dependent Hubbard \emph{U}. We
refer the static value of $W_{r}(0)$ as the Hubbard \emph{U} used in model Hamiltonians.

In the following sections, we compute matrix elements of $W_{r}$ in the
Wannier basis. The calculation starts with a conventional LDA electronic
structure obtained by the full-potential LMTO method. The polarizability is
computed using Kohn-Sham eigenfunctions and eigenvalues. 
It can be computed efficiently for an arbitrary  number of frequencies 
at essentially the cost of just one frequency \cite{miyake00}. 
Other technical details are found elsewhere 
\cite{aryasetiawan94,aryasetiawan00,kotani02,schilfgaarde06}. 
We use a $8\times8\times8$ k mesh for transition metals and a $4\times4\times4$ 
k mesh for SrVO$_{3}$
for the Brillouin zone integration. Optimization of Wannier functions is done
following the procedure in Ref.\cite{marzari97,souza01}.

\section{cRPA with the maximally localized Wannier function}

\begin{figure}[ptb]
\begin{center}
\includegraphics[width=90mm]{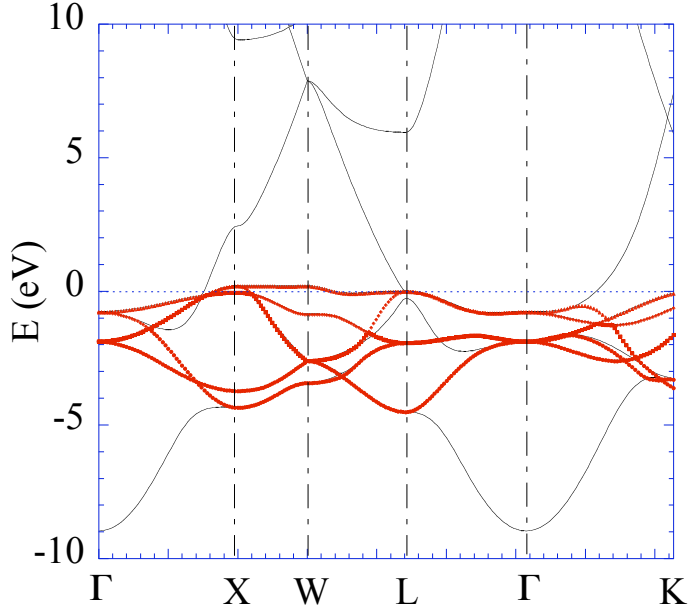}
\end{center}
\caption{Bandstructure of paramagnetic nickel in LDA (solid thin lines) and
projected bands (dotted thick lines). }%
\label{fig:band}%
\end{figure}

Let us start with paramagnetic nickel as an example. We first construct five
Wannier orbitals having strong 3$d$ character. Once $\mathcal{U}%
_{mn}(\mathbf{k})$ is determined on a k mesh, maximally localized Wannier
functions are obtained by Fourier transform, from which the Hamiltonian
$H_{mn}(\mathbf{R})=\langle\varphi_{m\mathbf{0}}|H|\varphi_{n\mathbf{R}%
}\rangle$ is reduced as well. By Fourier transforming $H_{mn}(\mathbf{R})$
back to k space and diagonalizing it \cite{souza01}, we can project out narrow
bands (Fig.\ref{fig:band}).

\begin{figure}[ptb]
\begin{center}
\includegraphics[width=90mm]{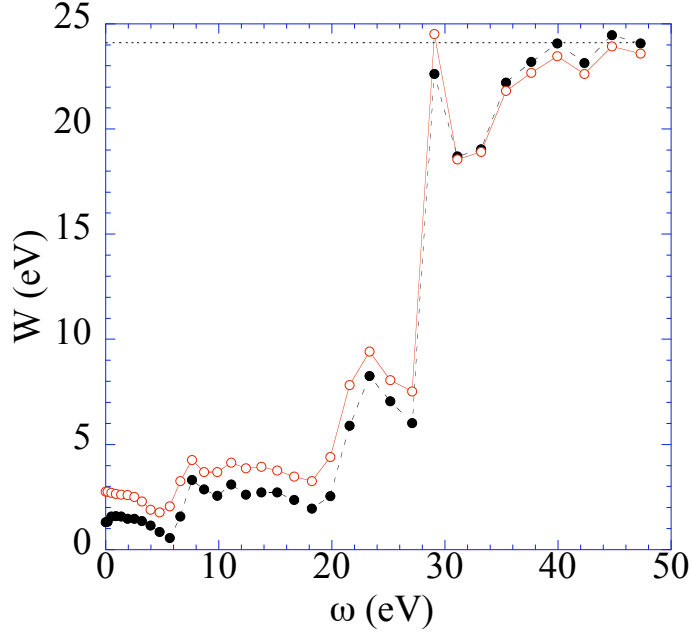}
\end{center}
\caption{ On-site screened Coulomb interaction of paramagnetic nickel in the
maximally localized Wannier basis as a function of frequency. Fully-screened
interaction (closed circles), result from cRPA (open circles), and bare
Coulomb interaction (dotted line). }%
\label{fig:wvsw}%
\end{figure}

The next step is to compute the screened Coulomb interaction $W_{r}%
(\mathbf{r}, \mathbf{r}^{\prime};\omega)$ in cRPA and take the matrix elements
in the maximally localized Wannier basis:%

\begin{equation}
W_{r}(n_{1},n_{2},n_{3},n_{4};\mathbf{R};\omega)\equiv\int\int\varphi
_{n_{1}\mathbf{0}}^{\ast}(\mathbf{r})\varphi_{n_{2}\mathbf{0}}(\mathbf{r}%
)W_{r}(\mathbf{r},\mathbf{r}^{\prime};\omega)\varphi_{n_{3}\mathbf{R}}^{\ast
}(\mathbf{r}^{\prime})\varphi_{n_{4}\mathbf{R}}(\mathbf{r}^{\prime}%
)d^{3}rd^{3}r^{\prime}\;. \label{Wr}%
\end{equation}
At this point it is worth pointing out that the effective screened interaction
$W_{r}(\mathbf{r},\mathbf{r}^{\prime};\omega)$
calculated using the cRPA method is completely independent of the choice of
basis functions. The matrix elements are of course dependent on the choice of
the orbitals $\varphi_{n\mathbf{0}}(\mathbf{r})$. The on-site diagonal
elements, $W_{r}(n,n,n,n,\mathbf{R=0};\omega)$ $(n=1,\cdots,5)$, are split by
crystal field effect, however the splitting is negligibly small. Figure
\ref{fig:wvsw} shows the average of the five, where fully screened Coulomb
interaction is also shown for comparison. $W_{r}$ is close to the bare Coulomb
value (dot-dashed line) at high energy, since screening effect is minor. As
the frequency decreases down to around 30 eV, the screening becomes effective
and $W_{r}$ decreases rapidly. At lower energy $W_{r}$ is weakly energy
dependent again and reaches 2.8 eV at $\omega=0$. These features are the same
as the previous calculations \cite{aryasetiawan04,aryasetiawan06}, though the
values are slightly smaller in the present result. The difference may be
ascribed to the difference in spacial extent of the orbitals $\varphi
_{n\mathbf{0}}(\mathbf{r})$. In the previous calculations, the effective
screened interaction $W_{r}(\mathbf{r},\mathbf{r}^{\prime};\omega)$ is
calculated within the LMTO-ASA scheme and the orbitals $\varphi_{n\mathbf{0}%
}(\mathbf{r})$ are taken to be the truncated partial waves, i.e., the heads of
the LMTO or the solutions of the Schr\"{o}dinger equation inside the atomic
sphere, rather than the LMTO\ basis. They are normalized and completely
confined to the atomic spheres. In the present calculations, $W_{r}%
(\mathbf{r},\mathbf{r}^{\prime};\omega)$ is calculated using the
full-potential LMTO (FP-LMTO) scheme but the orbitals $\varphi_{n\mathbf{0}}$
are the maximally localized Wannier functions, which have tail extending
outside the central cell. These orbitals are thus more delocalized and
consequently the matrix elements of $W_{r}$ are smaller. The difference in
$W_{r}(\mathbf{r},\mathbf{r}^{\prime};\omega)$ arising from the difference
between LMTO-ASA and FP-LMTO is probably less significant.

The above informations, projected band structure and on-site values of $W_{r}%
$, are key ingredients for constructing effective models. However, it is not
sensible to construct a Hubbard model by simply adding the former as the
kinetic term and the static value of the latter as the interaction term. In
fact, excitation spectra for that Hamiltonian do not reproduce the solution
of the original Hamiltonian. This is because (i) the kinetic term is
renormalized during the downfolding process, and (ii) $W_{r}$ is energy
dependent and long-ranged. However, there is an approximate way to construct a
Hubbard Hamiltonian with a static interaction \cite{aryasetiawan04}.

\begin{figure}[ptb]
\begin{center}
\includegraphics[width=90mm]{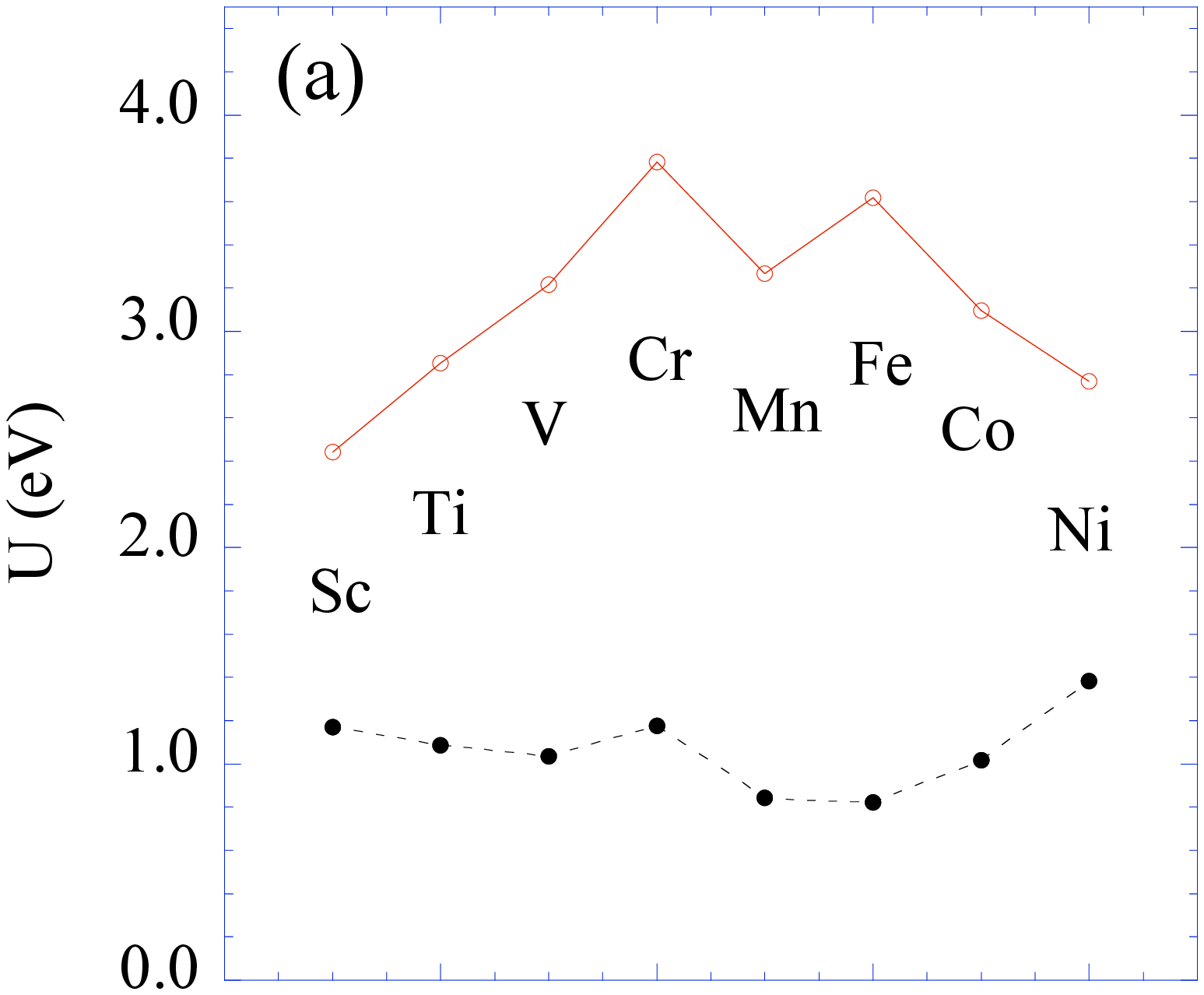} \includegraphics[width=90mm]{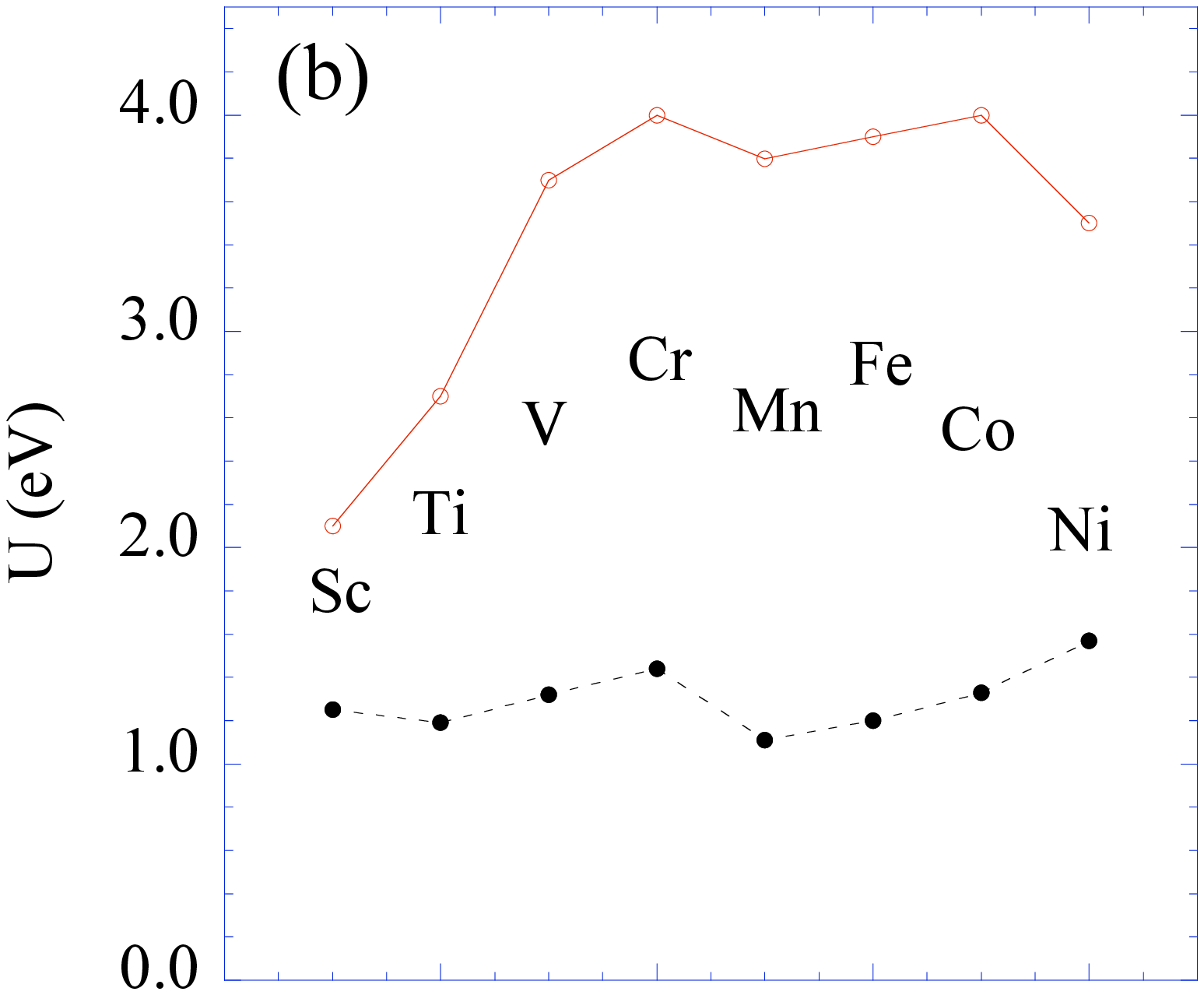}
\end{center}
\caption{ On-site screened Coulomb interaction $U$ for transition metals from
cRPA (open circles) and fully screened $U$ from RPA (closed circles). (a)
Results when the matrix elements of $U=W_{r}(\omega=0)$ are taken in the
maximally localized Wannier function basis, and (b) in the truncated partial
waves or the heads of the LMTO-ASA basis. }%
\label{fig:U}%
\end{figure}

Figure \ref{fig:U}(a) shows the diagonal element of the on-site $W_{r}$ in the
static limit ($U$) for a series of transition metals. Comparing the results
with those obtained from previous calculations presented in (b), the trend is
the same: As the atomic number increases, the $U$ increases in the early
transition metals, while it decreases in late transition metals. We also
computed $U$ in SrVO$_{3}$ (Table \ref{tab:svo}). In this system, there are
three $t_{2g}$ states near the Fermi level and they are isolated from other
bands. Thus, there is no ambiguity for dividing the space into $\{\psi_{d}\}$
and $\{\psi_{r}\}$. The value of $U$ is computed to be 3.0 eV, which is again
smaller than the previously calculated value of 3.5 eV. We emphasize again
that in the previous calculations the effective screened interaction $W_{r}$
is calculated within the LMTO-ASA scheme and the choice of the orbitals in
calculating the matrix elements of $W_{r}$ are truncated partial waves, which
are confined within the atomic sphere and thus more localized compared with
the Wannier orbitals used in the present calculations, leading to a larger
value of \emph{U}. The $U$ value obtained for SrVO$_{3}$ should be similar for
CaVO$_{3}$, LaTiO$_{3}$, and YTiO$_{3}$ perovskites.

\begin{table}[ptb]
\caption{ On-site Coulomb ($U$), exchange ($J$), and off-site Coulomb
($U^{\prime}$) energy in SrVO$_{3}$ obtained by cRPA. }%
\label{tab:svo}
\begin{ruledtabular}
\begin{tabular}{c c}
$U$ & 3.0 eV \\
$U^{\prime}$ & 0.45 eV \\
$J$ & 0.43 eV \\
\end{tabular}
\end{ruledtabular}
\end{table}

\begin{figure}[ptb]
\begin{center}
\includegraphics[width=90mm]{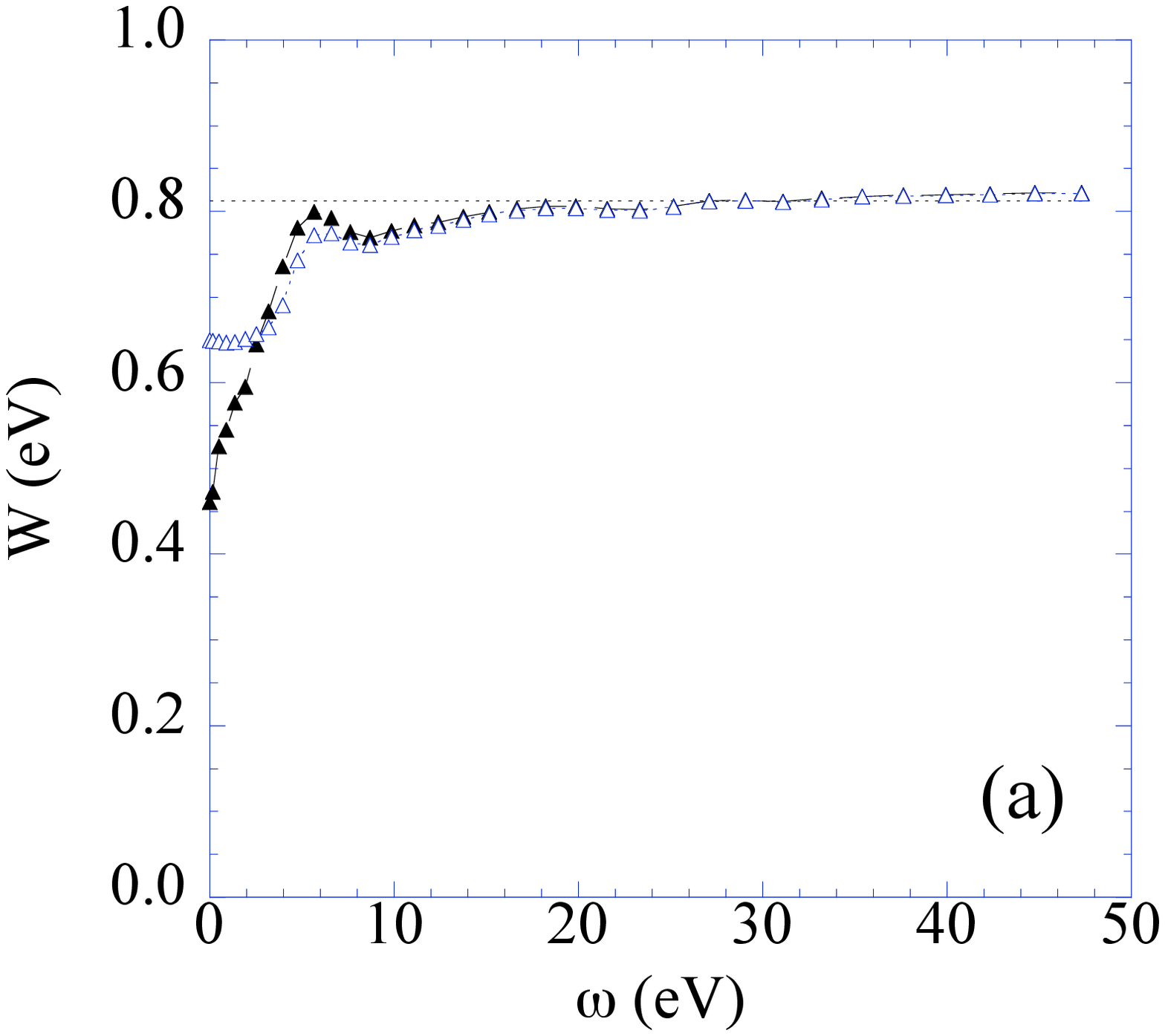} \includegraphics[width=90mm]{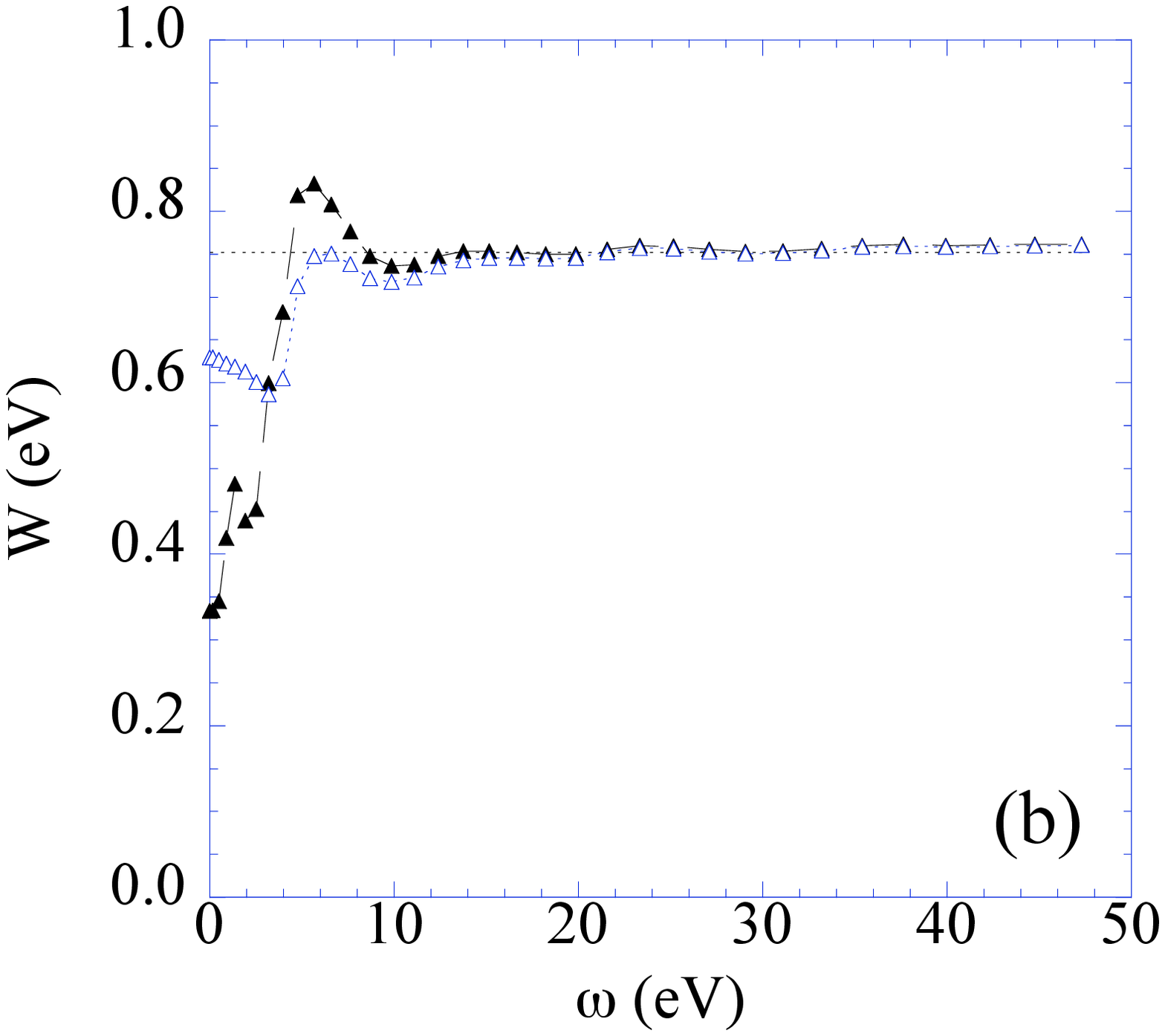}
\includegraphics[width=90mm]{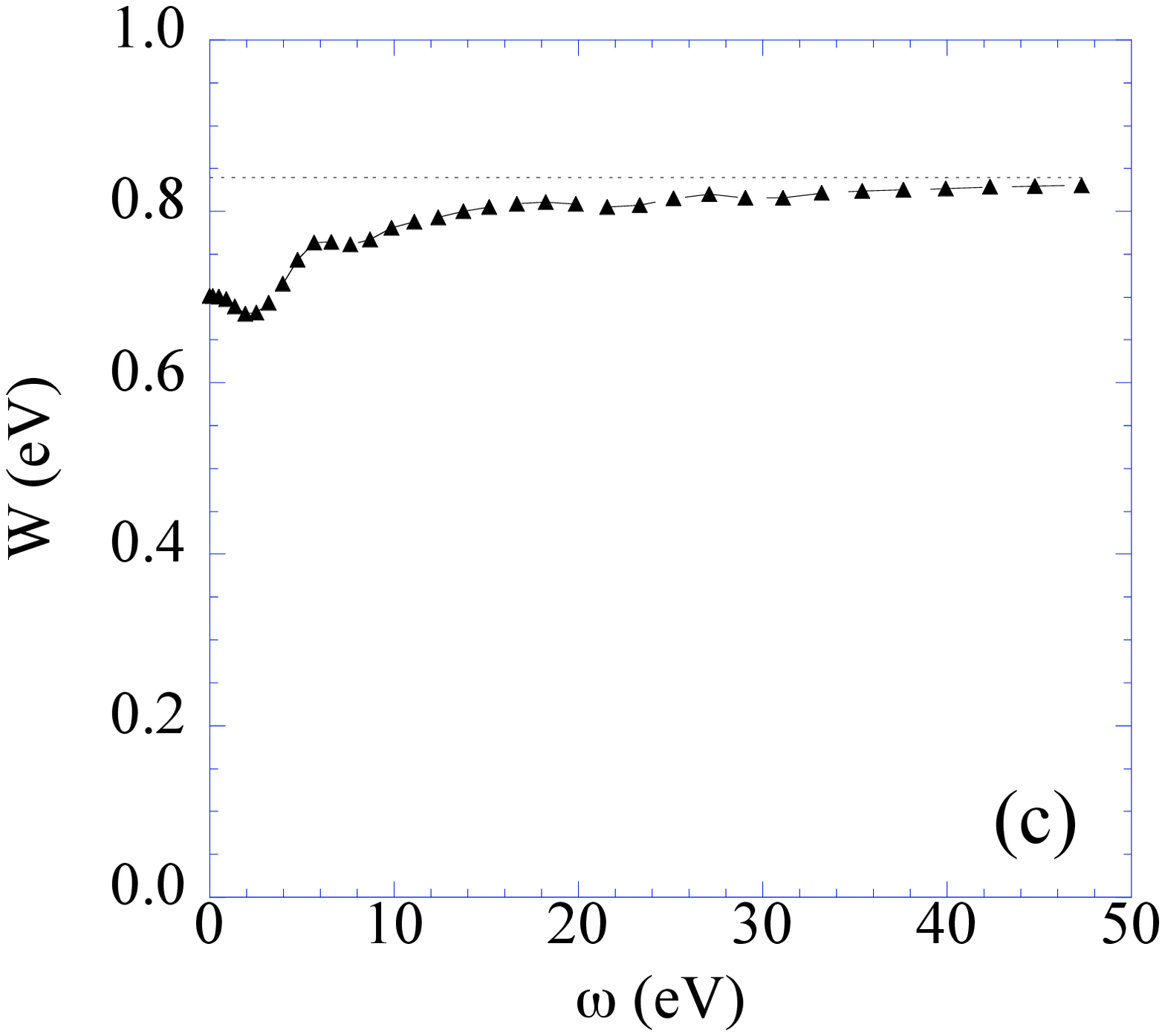}
\end{center}
\caption{ On-site exchange interaction from fully screened interaction (closed
triangles), cRPA (open triangles), and from bare interaction (dotted line) in
(a) Ni, (b) Fe, and (c) Cu.}%
\label{fig:jvsw}%
\end{figure}

\begin{figure}[ptb]
\begin{center}
\includegraphics[width=90mm]{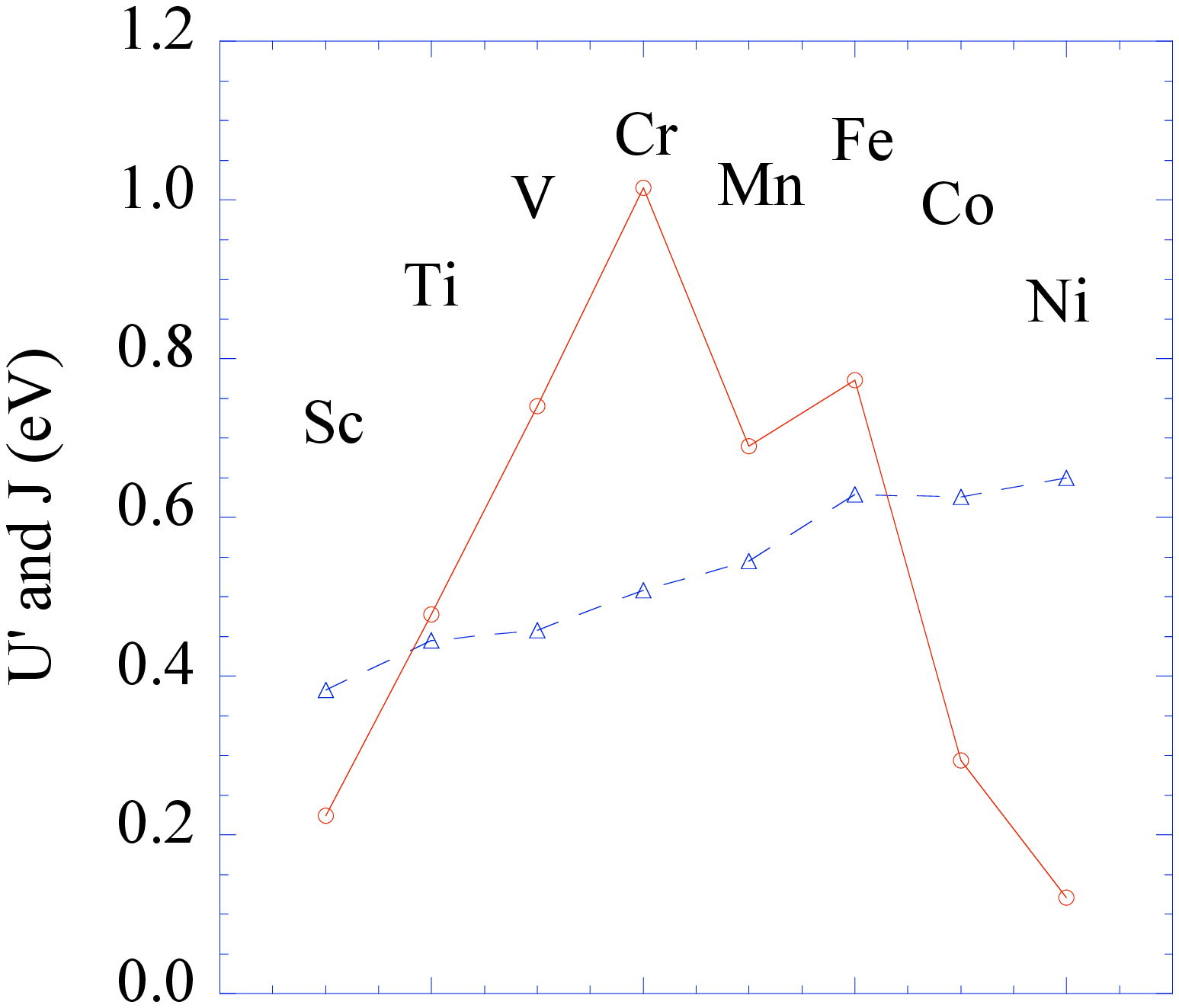}
\end{center}
\caption{ On-site screened exchange interaction $J$ (triangles) and off-site
screened Coulomb interaction $U^{\prime}$ (open circles). }%
\label{fig:J}%
\end{figure}

The off-diagonal (exchange) elements are also important quantities. In
Fig.\ref{fig:jvsw}(a), $\langle W_{r}(n,m,m,n,\mathbf{R=0};\omega
)\rangle_{n\neq m}$ in Ni is shown as a function of frequency, where average
is taken over $n$ and $m$. In contrast with the Coulomb term, the exchange
term is weakly energy dependent and does not show significant change at around
30 eV. We may understand this behavior as follows. $\int d^{3}r^{\prime}%
W_{r}(\mathbf{r},\mathbf{r}^{\prime};\omega)\varphi_{n_{3}\mathbf{R}}^{\ast
}(\mathbf{r}^{\prime})\varphi_{n_{4}\mathbf{R}}(\mathbf{r}^{\prime})$ is a
screened potential of a charge density $\rho(\mathbf{r}^{\prime}%
)=\varphi_{n_{3}\mathbf{R}}^{\ast}(\mathbf{r}^{\prime})\varphi_{n_{4}%
\mathbf{R}}(\mathbf{r}^{\prime})$. In accordance with known observation, the
potential arising from an exchange charge density ($n_{3}\neq n_{4}$) is not
well screened because it has no mono pole (zero spherical average) in contrast
to the case of $n_{3}=n_{4}$. At the onset of the plasmon excitation at around
30 eV, a perturbing charge is highly screened and the screening is
electron-gas-like, which is highly effective for a mono pole. At lower energy
(5-6 eV), however, atomic-like screening in the form of 3d-3d transitions as
well as 3d-4p takes place, which is relatively effective in screening a
multipole charge distribution, resulting in a significant decrease of $J$. As
can be seen in the case of Fe and Ni, $J$ is reduced considerably when 3d-3d
screening arising from $P_{d}$ is included whereas in Cu, where there are
essentially no 3d-3d transitions, since the 3d band is fully occupied, the
main screening channels come from 3d-4p transitions and $J$ varies less
strongly compared to those of Ni and Fe for the fully screened case. Figure
\ref{fig:J} shows the static value of the exchange term, $J$. We find that $J$
does not depend on the element significantly and its value is around 0.5 eV.

\begin{figure}[ptb]
\begin{center}
\includegraphics[width=90mm]{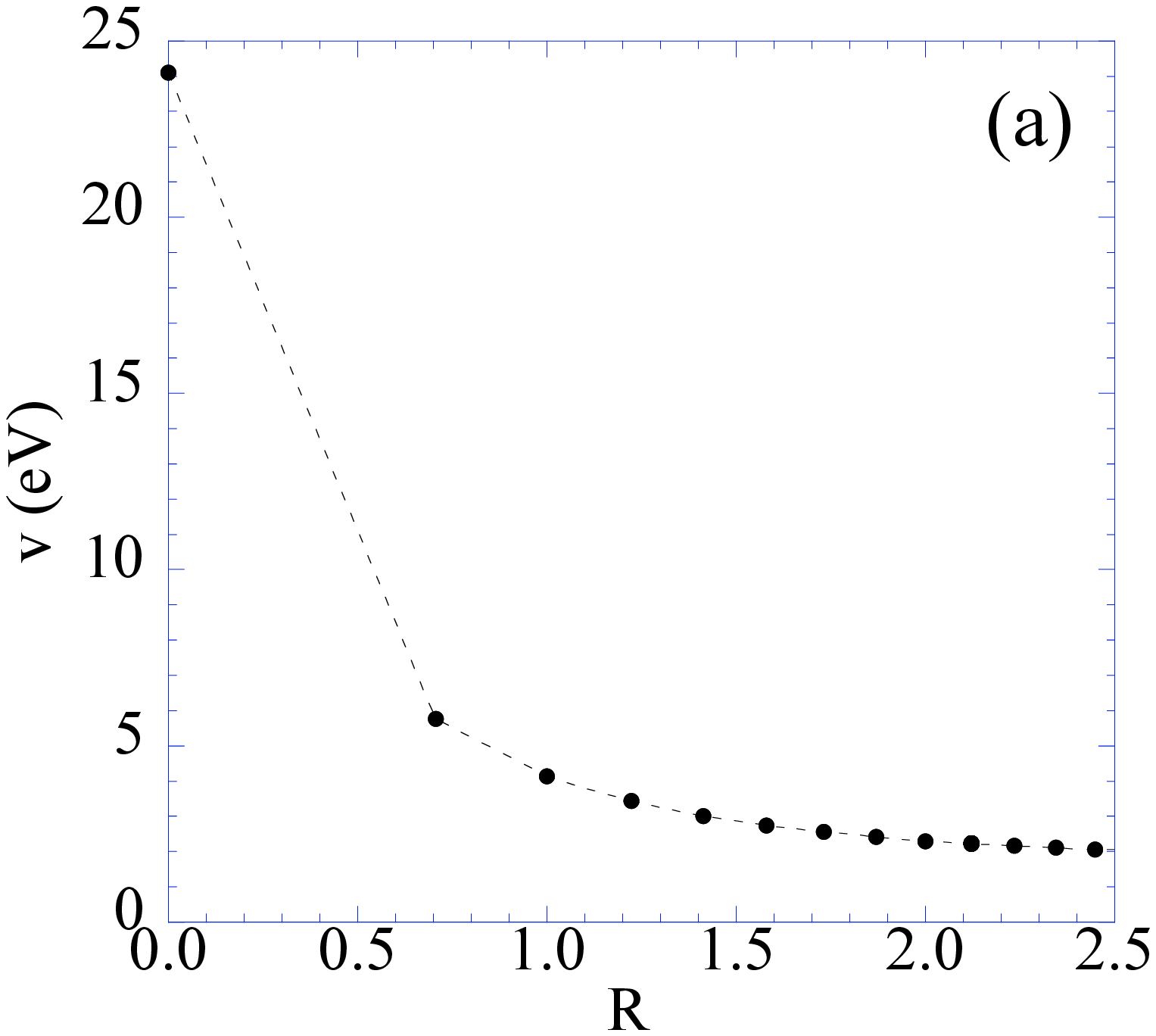} \includegraphics[width=90mm]{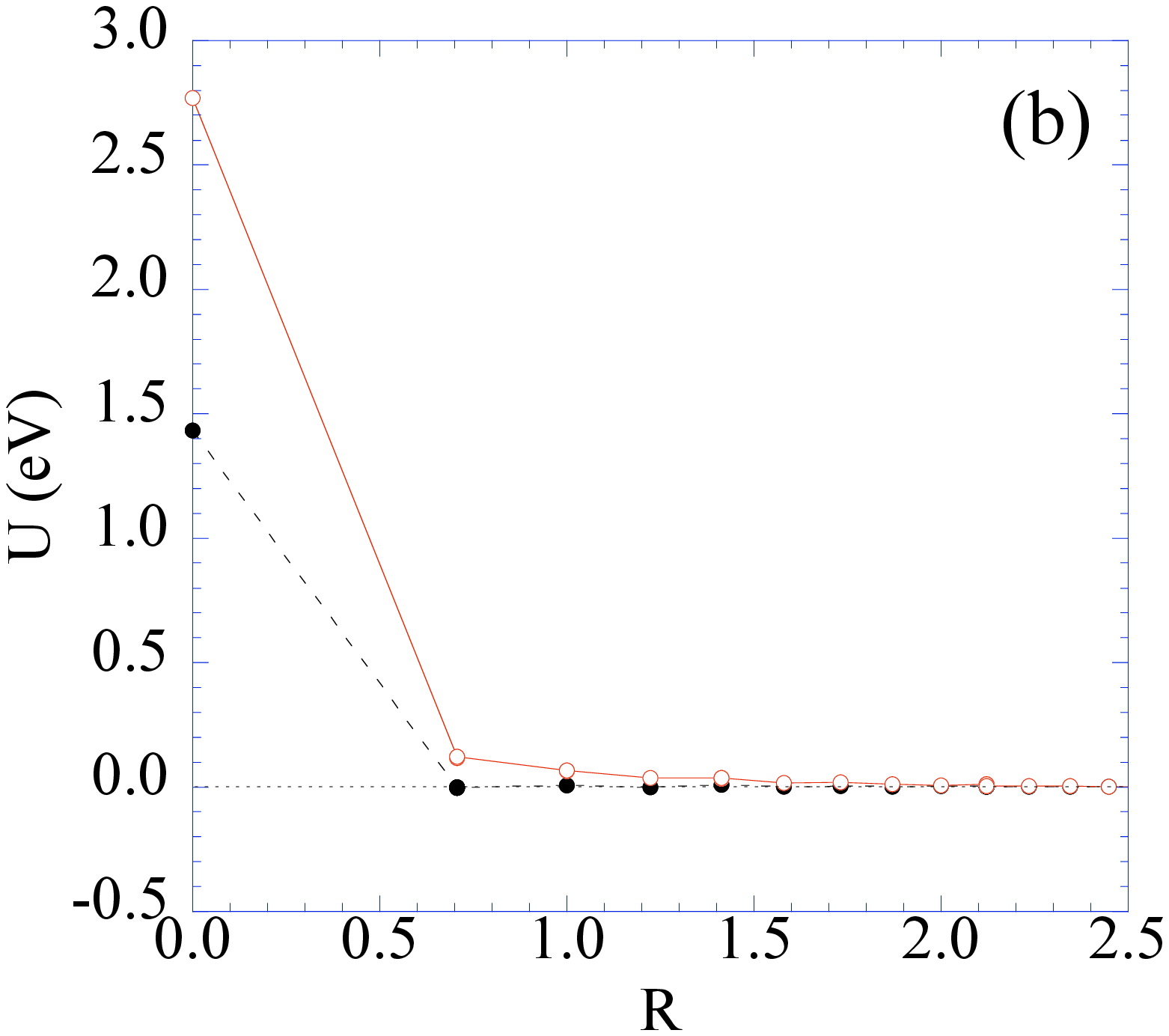}
\end{center}
\caption{ Diagonal elements of Coulomb interaction of Ni as a function of $R=|
\mathbf{R} |$ (in units of lattice constant). (a) bare Coulomb interaction,
and (b) Screened Coulomb interactions in cRPA (open circles) and in RPA
(closed circles). }%
\label{fig:ur}%
\end{figure}

Another important information is non-locality of the interactions. The bare
Coulomb interaction $v$ as a function of $R=|\mathbf{R}|$ is long ranged
(Fig.\ref{fig:ur}(a)), and $v$ between the nearest neighbor cells is about 1/4
of the on-site value.
On the other hand, the screened interaction shows much faster damping and the
value at the nearest neighbor ($U^{\prime}$) is 0.1 eV, which is much smaller
than the on-site value of $U$=2.7 eV. The values of $U^{\prime}$ for other
transition metals are shown by open circles in Fig.\ref{fig:J}.

The approximately parabolic variation on both $U$ and $U^{\prime}$ across the
series may be qualitatively understood in terms of 3d band filling. The
largest polarization inside the 3d band ($P_{d}$) corresponds roughly to
half-filling. Since this is eliminated when calculating $U$ and $U^{\prime}$,
they peak around the middle of the series. This is on contrast to the fully
screened interaction $W$ which is almost a constant across the series. As
discussed in \cite{aryasetiawan04} since the screening is metallic it does not
depend much on the element: there are always enough electrons to screen a
perturbing charge. Moreover, since the screened interaction is rather
localized, it is not sensitive to the extent of the orbitals used in taking
the matrix elements in Eq.(\ref{Wr}).

The $U^{\prime}$ is larger in some elements. For example, $U^{\prime}$ of Cr
is as large as 1.0 eV, which would not be negligible. In such cases, one would
wish to reconstruct Wannier function so that off-site interaction is as small
as possible. This possibility is discussed in the next section, where a
procedure to maximize the on-site $U$ is derived and applied.

\section{Maximizing the on-site \emph{U} parameter}

\subsection{Formulation}

We follow closely the method in Ref. \cite{edmiston63} and apply it to the
case of periodic crystals. We use the convention that repeated indices are
summed. Let us define a unitary transformation%

\begin{align}
\chi_{n\mathbf{R}}  &  =\varphi_{n\mathbf{R}}+\delta\varphi_{n\mathbf{R}%
}\nonumber\\
&  =\varphi_{n^{\prime}\mathbf{R}^{\prime}}T_{n^{\prime}\mathbf{R}^{\prime
},n\mathbf{R}}\;, \label{chi0}%
\end{align}

\begin{equation}
T_{n^{\prime}\mathbf{R}^{\prime},n\mathbf{R}}=\delta_{nn^{\prime}}%
\delta_{\mathbf{RR}^{\prime}}+\tau_{n^{\prime}\mathbf{R}^{\prime},n\mathbf{R}%
}\;.
\end{equation}
From the unitarity of $T$ one has, to first order%

\begin{equation}
\delta\varphi_{n\mathbf{R}}=\varphi_{n^{\prime}\mathbf{R}^{\prime}}%
\tau_{n^{\prime}\mathbf{R}^{\prime},n\mathbf{R}},\ \ \tau_{n^{\prime
}\mathbf{R}^{\prime},n\mathbf{R}}^{+}+\tau_{n^{\prime}\mathbf{R}^{\prime
},n\mathbf{R}}=0 \;.
\end{equation}
Consider a change in \emph{U} to first order in $\delta\varphi_{n\mathbf{R}}$.%

\begin{align}
U  &  =\sum_{n\mathbf{R}}\left\langle \chi_{n\mathbf{R}}^{\ast}\chi
_{n\mathbf{R}}|U|\chi_{n\mathbf{R}}^{\ast}\chi_{n\mathbf{R}}\right\rangle \;\\
\delta U  &  =\left\langle \delta\varphi_{n\mathbf{R}}^{\ast}\varphi
_{n\mathbf{R}}|U|\varphi_{n\mathbf{R}}^{\ast}\varphi_{n\mathbf{R}%
}\right\rangle +\left\langle \varphi_{n\mathbf{R}}^{\ast}\delta\varphi
_{n\mathbf{R}}|U|\varphi_{n\mathbf{R}}^{\ast}\varphi_{n\mathbf{R}%
}\right\rangle \nonumber\\
&  +\left\langle \varphi_{n\mathbf{R}}^{\ast}\varphi_{n\mathbf{R}}%
|U|\delta\varphi_{n\mathbf{R}}^{\ast}\varphi_{n\mathbf{R}}\right\rangle
+\left\langle \varphi_{n\mathbf{R}}^{\ast}\varphi_{n\mathbf{R}}|U|\varphi
_{n\mathbf{R}}^{\ast}\delta\varphi_{n\mathbf{R}}\right\rangle \nonumber\\
&  =\left\langle \varphi_{n^{\prime}\mathbf{R}^{\prime}}^{\ast}\varphi
_{n\mathbf{R}}|U|\varphi_{n\mathbf{R}}^{\ast}\varphi_{n\mathbf{R}%
}\right\rangle \tau_{n^{\prime}\mathbf{R}^{\prime}\mathbf{,}n\mathbf{R}}%
^{\ast}+\left\langle \varphi_{n\mathbf{R}}^{\ast}\varphi_{n^{\prime}%
\mathbf{R}^{\prime}}|U|\varphi_{n\mathbf{R}}^{\ast}\varphi_{n\mathbf{R}%
}\right\rangle \tau_{n^{\prime}\mathbf{R}^{\prime}\mathbf{,}n\mathbf{R}%
}\nonumber\\
&  +\left\langle \varphi_{n\mathbf{R}}^{\ast}\varphi_{n\mathbf{R}}%
|U|\varphi_{n^{\prime}\mathbf{R}^{\prime}}^{\ast}\varphi_{n\mathbf{R}%
}\right\rangle \tau_{n^{\prime}\mathbf{R}^{\prime}\mathbf{,}n\mathbf{R}}%
^{\ast}+\left\langle \varphi_{n\mathbf{R}}^{\ast}\varphi_{n\mathbf{R}%
}|U|\varphi_{n\mathbf{R}}^{\ast}\varphi_{n^{\prime}\mathbf{R}^{\prime}%
}\right\rangle \tau_{n^{\prime}\mathbf{R}^{\prime}\mathbf{,}n\mathbf{R}%
}\nonumber\\
&  =2\left\langle \varphi_{n^{\prime}\mathbf{R}^{\prime}}^{\ast}%
\varphi_{n\mathbf{R}}|U|\varphi_{n\mathbf{R}}^{\ast}\varphi_{n\mathbf{R}%
}\right\rangle \tau_{n^{\prime}\mathbf{R}^{\prime}\mathbf{,}n\mathbf{R}}%
^{\ast}+2\left\langle \varphi_{n\mathbf{R}}^{\ast}\varphi_{n^{\prime
}\mathbf{R}^{\prime}}|U|\varphi_{n\mathbf{R}}^{\ast}\varphi_{n\mathbf{R}%
}\right\rangle \tau_{n^{\prime}\mathbf{R}^{\prime}\mathbf{,}n\mathbf{R}}\;.
\end{align}
This provides an expression for the change of \emph{U} as a function of
independent parameters $\tau_{n^{\prime}\mathbf{R}^{\prime}\mathbf{,}%
n\mathbf{R}}^{\ast}$ and $\tau_{n^{\prime}\mathbf{R}^{\prime}\mathbf{,}%
n\mathbf{R}}$. Using $\tau_{n^{\prime}\mathbf{R}^{\prime}\mathbf{,}%
n\mathbf{R}}^{\ast}+\tau_{n\mathbf{R,}n^{\prime}\mathbf{R}^{\prime}}=0$ one
can rewrite this expression as follows.%

\begin{align}
\delta U  &  =-2\left\langle \varphi_{n^{\prime}\mathbf{R}^{\prime}}^{\ast
}\varphi_{n\mathbf{R}}|U|\varphi_{n\mathbf{R}}^{\ast}\varphi_{n\mathbf{R}%
}\right\rangle \tau_{n\mathbf{R,}n^{\prime}\mathbf{R}^{\prime}}+2\left\langle
\varphi_{n\mathbf{R}}^{\ast}\varphi_{n^{\prime}\mathbf{R}^{\prime}}%
|U|\varphi_{n\mathbf{R}}^{\ast}\varphi_{n\mathbf{R}}\right\rangle
\tau_{n^{\prime}\mathbf{R}^{\prime},n\mathbf{R}}\nonumber\\
&  =2\left[  \left\langle \varphi_{n\mathbf{R}}^{\ast}\varphi_{n^{\prime
}\mathbf{R}^{\prime}}|U|\varphi_{n\mathbf{R}}^{\ast}\varphi_{n\mathbf{R}%
}\right\rangle -\left\langle \varphi_{n\mathbf{R}}^{\ast}\varphi_{n^{\prime
}\mathbf{R}^{\prime}}|U|\varphi_{n^{\prime}\mathbf{R}^{\prime}}^{\ast}%
\varphi_{n^{\prime}\mathbf{R}^{\prime}}\right\rangle \right]  \tau_{n^{\prime
}\mathbf{R}^{\prime},n\mathbf{R}}\nonumber\\
&  =2F_{n\mathbf{R,}n^{\prime}\mathbf{R}^{\prime}}^{+}\tau_{n^{\prime
}\mathbf{R}^{\prime},n\mathbf{R}} \;,
\end{align}
where we have defined an anti-Hermitian matrix ($F^{+}=-F$)%

\begin{equation}
F_{n\mathbf{R,}n^{\prime}\mathbf{R}^{\prime}}^{+}=\left\langle \varphi
_{n\mathbf{R}}^{\ast}\varphi_{n^{\prime}\mathbf{R}^{\prime}}|U|\varphi
_{n\mathbf{R}}^{\ast}\varphi_{n\mathbf{R}}\right\rangle -\left\langle
\varphi_{n\mathbf{R}}^{\ast}\varphi_{n^{\prime}\mathbf{R}^{\prime}}%
|U|\varphi_{n^{\prime}\mathbf{R}^{\prime}}^{\ast}\varphi_{n^{\prime}%
\mathbf{R}^{\prime}}\right\rangle =F_{n^{\prime}\mathbf{R}^{\prime
},n\mathbf{R}}^{\ast} \;. \label{F}%
\end{equation}
We now choose%

\begin{equation}
\tau_{n^{\prime}\mathbf{R}^{\prime},n\mathbf{R}}=\varepsilon F_{n^{\prime
}\mathbf{R}^{\prime},n\mathbf{R}}\;,
\end{equation}
which ensures that%

\begin{equation}
\delta U(\varepsilon)=2\varepsilon F_{n\mathbf{R,}n^{\prime}\mathbf{R}%
^{\prime}}^{+}F_{n^{\prime}\mathbf{R}^{\prime},n\mathbf{R}}\geqq0 \;.
\label{dU}%
\end{equation}
The procedure is then the following. Construct the matrix%
\begin{equation}
T=e^{\varepsilon F} \;,
\end{equation}
which is unitary because $F$ is anti-Hermitian. To calculate $T$ we
diagonalize $F$ with eigenvectors $\,\left\langle n\mathbf{R|}\alpha
\right\rangle $ and eigenvalues $f_{\alpha}$:%

\begin{equation}
T_{n^{\prime}\mathbf{R}^{\prime},n\mathbf{R}}(\varepsilon)=\left\langle
n^{\prime}\mathbf{R}^{\prime}|\alpha\right\rangle e^{\varepsilon f_{\alpha}%
}\left\langle \alpha|n\mathbf{R}\right\rangle \;.
\end{equation}
One now obtains a new basis and calculates a new $U$ as a function of
$\varepsilon$.%

\begin{align}
U(\varepsilon)  &  =\left\langle \chi_{n\mathbf{R}}^{\ast}\chi_{n\mathbf{R}%
}|U|\chi_{n\mathbf{R}}^{\ast}\chi_{n\mathbf{R}}\right\rangle \nonumber\\
&  =\left\langle \varphi_{i\mathbf{R}_{1}}^{\ast}\varphi_{j\mathbf{R}_{2}%
}|U|\varphi_{k\mathbf{R}_{3}}^{\ast}\varphi_{l\mathbf{R}_{4}}\right\rangle
T_{i\mathbf{R}_{1},n\mathbf{R}}^{\ast}T_{j\mathbf{R}_{2},n\mathbf{R}%
}T_{k\mathbf{R}_{3},n\mathbf{R}}^{\ast}T_{l\mathbf{R}_{4},n\mathbf{R}} \;.
\end{align}
One varies $\varepsilon$ until $U(\varepsilon)$ reaches a maximum and chooses
that new basis set $\chi_{n\mathbf{R}}$ that maximizes $U(\varepsilon)$. The
procedure is then repeated until convergence is achieved. In practice, we
solve for the eigenvectors and eigenvalues of $iF$, which is Hermitian. The
eigenvalues of $F$ are then given by $-i$ times the eigenvalues of $iF$. The
starting orbitals $\varphi_{n\mathbf{R}}$ are chosen to be the maximally
localized Wannier orbitals but other choices are also possible.

The key quantity in the present formulation is the anti-Hermitian matrix $F$,
which is defined with respect to a given cluster. For finite systems such as
molecules it is clear how to apply the above formulation \cite{edmiston63}.
One simply constructs the matrix $F$ from the definition in Eq.(\ref{F}),
where $\mathbf{R,R}^{\prime}$ run over the sites in the molecule. It is not
however immediately clear how to apply the method to periodic crystals. For
this purpose we define a cluster or supercell around the unit cell (site) at
the origin. A new Wannier orbital centered at the origin is constructed as a
linear combination of orbitals centered on the sites in the cluster as in Eq.(\ref{chi0}). 
For simplicity but without loss of generality let us consider
the case of one orbital per site (unit cell).\ First we note that
$F_{\mathbf{R}^{\prime},\mathbf{R}}$ depends only on the relative distance,
$F_{\mathbf{R}^{\prime},\mathbf{R}}=F_{\mathbf{R}^{\prime}-\mathbf{R,0}}$. For
$\mathbf{R=0}$, we obtain the first column $F_{\mathbf{R}^{\prime},\mathbf{0}%
}$ according to the definition in Eq.(\ref{F}). We now move to another site
$\mathbf{R}$ in the cluster and construct a new Wannier orbital centered on
this site $\mathbf{R}$ as a linear combination of orbitals centered on the
same cluster sites but shifted by $\mathbf{R}$ with respect to the cluster
centered at the origin. This is to ensure that the Wannier orbitals so
constructed will be independent of the sites. The cluster centered at
$\mathbf{R}$, however, has some of its sites outside the original cluster
centered at the origin. But there is a correspondence between those sites
outside and those sites inside the original cluster: sites connected by
superlattice vectors are equivalent. This allows us to construct
$F_{\mathbf{R}^{\prime},\mathbf{R}}$ from $F_{\mathbf{R}^{\prime},\mathbf{0}}$
in the following way. We search for a superlattice translational vector
$\mathbf{T}$ such that%

\begin{equation}
\mathbf{R}_{0}=\mathbf{R}^{\prime}-\mathbf{R-T} \;,
\end{equation}
is a site in the original cluster centered at the origin. Then the column
$F_{\mathbf{R}^{\prime},\mathbf{R}}$ is given by%

\begin{equation}
F_{\mathbf{R}^{\prime},\mathbf{R}}=F_{\mathbf{R}_{0},\mathbf{0}}\;.
\end{equation}
A given column of $F$ consists of permuted elements of the other columns.

\subsection{Results}

Using the procedure described in the previous section we have constructed
Wannier orbitals by maximizing \emph{U} for the 3d transition metals. The
cluster consists of the nearest and next nearest neighbor sites. As can be
seen in Table \ref{tab:maxu} the resulting \emph{U} values are remarkably
close to the values calculated using the maximally localized Wannier orbitals.
The results do not change in any significant way when only nearest neighbors
are included in the cluster. We confirm that the change in \emph{U} compares
favorably with the estimated value in Eq.(\ref{dU}). Indeed we have checked
that the coefficients of expansion $T_{n^{\prime}\mathbf{R}^{\prime
},n\mathbf{0}}$ is essentially unity when $\mathbf{R}^{\prime}=\mathbf{0}%
,n^{\prime}=n$ and zero otherwise. This implies that the maximally localized
Wannier orbitals at the same time to a very good approximation maximize the
on-site \emph{U} and form a good basis for the construction of low-energy
model Hamiltonians such as the Hubbard model.

To convince ourselves that our procedure is sound, we have performed the
following calculations. We construct a maximally localized Wannier orbital
corresponding to \emph{xy} orbital of the $t_{2g}~$symmetry of SrVO$_{3}$ and
calculate \emph{U}. We also construct Wannier orbitals that are deliberately
delocalized but span the same Hilbert space as that of the maximally localized
ones and calculate the corresponding \emph{U}. A 2$\times$2$\times$2 k mesh is
used in the calculations. The results are shown in Table \ref{tab:maxusvo}. As
expected, the value of \emph{U} corresponding to the delocalized orbital is
considerably smaller than that corresponding to the maximally localized one.
We now construct a Wannier orbital by forming linear combinations of both the
maximally localized orbitals as well as the delocalized ones and maximize
\emph{U}. 
The orbitals are centered on sites shown in Table \ref{tab:dist}, 
The maximum \emph{U}'s\emph{ }calculated from the unitary
transformation of the maximally localized orbitals and the delocalized
orbitals consistently agree with each other, as they should since the
maximally localized orbitals and the delocalized orbitals span the same
Hilbert space. As in the case of the 3d transition metals, the maximized
\emph{U} is essentially identical to the value corresponding to the maximally
localized Wannier orbital. In Table \ref{tab:dist} we show the distribution of
weight, $|T_{\mathbf{R},\mathbf{0}}|^{2}$, of the orbital that maximizes
\emph{U} formed by a linear combination of the delocalized orbitals. The
result indicates that the original maximally localized Wannier orbital
centered at (1 1 1) is the only one that has a significant weight at the
origin. The corresponding weights for the maximally localized orbitals are
almost unity when $\mathbf{R=0}$ and zero otherwise.

\begin{table}[ptb]
\caption{The Hubbard $U$ calculated by maximizing the on-site $U$ compared
with the values obtained using the maximally localized Wannier orbitals. }%
\label{tab:maxu}
\begin{ruledtabular}
\begin{tabular}{l c c c}
& $U_0$ (eV) & Max. $U$ (eV) & $\delta U$ (eV)\\
\hline
Sc & 2.444119 & 2.444175 & 5.5E-005 \\
Ti & 2.853722 & 2.853747 & 2.6E-005 \\
V  & 3.216846 & 3.216978 & 1.3E-004 \\
Cr & 3.781924 & 3.782087 & 1.6E-004 \\
Mn & 3.268736 & 3.268788 & 5.2E-005 \\
Fe & 3.619025 & 3.619095 & 7.0E-005 \\
Co & 3.097218 & 3.097302 & 8.3E-005 \\
Ni & 2.769907 & 2.769935 & 2.7E-005 \\
\end{tabular}
\end{ruledtabular}
\end{table}

\begin{table}[ptb]
\caption{The Hubbard $U$ calculated by maximizing the on-site $U$ (Max. $U$)
for SrVO$_{3}$. Starting from the maximally localized Wannier orbitals and
delocalized orbitals consistently give the same maximum value of $U$. }%
\label{tab:maxusvo}
\begin{ruledtabular}
\begin{tabular}{l c c}
& $U_0$ (eV) & Max. $U$ (eV) \\
\hline
localized   & 3.3808733554 & 3.3808733554  \\
delocalized & 3.0292927908 & 3.3808733554  \\
\end{tabular}
\end{ruledtabular}
\end{table}

\begin{table}[ptb]
\caption{The distribution of weight of the Wannier orbitals that maximize $U$
when starting from delocalized orbitals for SrVO$_{3}$ }%
\label{tab:dist}
\begin{ruledtabular}
\begin{tabular}{c c}
Site & Weight \\
\hline
(0 0 0) & 0.939 \\
(0 0 1) & 4.1E-008 \\
(0 1 0) & 4.4E-007 \\
(0 1 1) & 4.8E-006 \\
(1 0 0) & 3.1E-007 \\
(1 0 1) & 6.8E-006 \\
(1 1 0) & 6.3E-007 \\
(1 1 1) & 0.061 \\
\end{tabular}
\end{ruledtabular}
\end{table}

It is remarkable that for the cases considered in the present work the Wannier
orbitals constructed by maximizing the on-site \emph{U} are almost identical
with those of maximally localized Wannier orbitals. Since the screened
interaction $W_{r}$ is deep around the Wannier center, it is reasonable to
expect that the maximally localized Wannier orbitals also yield a large value
of \emph{U }close to the maximum value. However, the extreme closeness to the
maximum value is rather unexpected. We have also performed the same
calculations by maximizing the bare Coulomb interaction and found very similar results.

\section{Real-space approach to maximally localized Wannier orbitals}

Finally we propose a real-space approach of constructing maximally localized
Wannier orbitals. We construct a unitary transformation on the Wannier
orbitals in real space and minimize%

\begin{equation}
\Omega=\sum_{\alpha}\left[  \left\langle r^{2}\right\rangle _{\alpha
}-\mathbf{\bar{r}}_{\alpha}^{2}\right]  \;.
\end{equation}
We use a combined notation $\alpha=(\mathbf{R}n)$ and the sum is restricted
over the sites in a cluster or supercell. As in the case of maximizing
\emph{U} consider a small variation%

\begin{equation}
\chi_{\alpha}=\varphi_{\alpha}+\delta\varphi_{\alpha}\;,
\end{equation}

\begin{equation}
\delta\varphi_{\alpha}=\varphi_{\beta}\tau_{\beta\alpha},\ \ \tau_{\alpha
\beta}^{\ast}+\tau_{\beta\alpha}=0\;,
\end{equation}

\begin{align}
\Omega &  =\sum_{\alpha}\left[  \left\langle r^{2}\right\rangle _{\alpha
}-\mathbf{\bar{r}}_{\alpha}^{2}\right] \nonumber\\
&  =\sum_{\alpha}\left[  \left\langle \chi_{\alpha}|r_{\alpha}^{2}%
|\chi_{\alpha}\right\rangle -\left\langle \chi_{\alpha}|\mathbf{r}_{\alpha
}|\chi_{\alpha}\right\rangle ^{2}\right] \nonumber\\
&  =\sum_{\alpha}\left[  \left\langle \varphi_{\alpha}+\delta\varphi_{\alpha
}|r_{\alpha}^{2}|\varphi_{\alpha}+\delta\varphi_{\alpha}\right\rangle
-\left\langle \varphi_{\alpha}+\delta\varphi_{\alpha}|\mathbf{r}_{\alpha
}|\varphi_{\alpha}+\delta\varphi_{\alpha}\right\rangle ^{2}\right]  \;,
\end{align}
$\mathbf{r}_{\alpha}$ means that the position is measured with respect to the
origin at site $\alpha$. The change in $\Omega$ to first order in
$\delta\varphi$ is%

\begin{align}
\delta\Omega &  =\sum_{\alpha}\left[  \left\langle \delta\varphi_{\alpha
}|r_{\alpha}^{2}|\varphi_{\alpha}\right\rangle +\left\langle \varphi_{\alpha
}|r_{\alpha}^{2}|\delta\varphi_{a}\right\rangle -2\left\langle \varphi
_{\alpha}|\mathbf{r}_{\alpha}|\varphi_{\alpha}\right\rangle \cdot\left\{
\left\langle \delta\varphi_{\alpha}|\mathbf{r}_{\alpha}|\varphi_{\alpha
}\right\rangle +\left\langle \varphi_{\alpha}|\mathbf{r}_{\alpha}%
|\delta\varphi_{\alpha}\right\rangle \right\}  \right] \nonumber\\
&  =\sum_{\alpha\beta}\left[  \left\langle \varphi_{\beta}|r_{\alpha}%
^{2}|\varphi_{\alpha}\right\rangle \tau_{\beta\alpha}^{\ast}+\left\langle
\varphi_{\alpha}|r_{\alpha}^{2}|\varphi_{\beta}\right\rangle \tau_{\beta
\alpha}-2\left\langle \varphi_{\alpha}|\mathbf{r}_{\alpha}|\varphi_{\alpha
}\right\rangle \cdot\left\{  \left\langle \varphi_{\beta}|\mathbf{r}_{\alpha
}|\varphi_{\alpha}\right\rangle \tau_{\beta\alpha}^{\ast}+\left\langle
\varphi_{\alpha}|\mathbf{r}_{\alpha}|\varphi_{\beta}\right\rangle \tau
_{\beta\alpha}\right\}  \right] \nonumber\\
&  =\sum_{\alpha\beta}\left[  \left\{  \left\langle \varphi_{\alpha}%
|r_{\alpha}^{2}|\varphi_{\beta}\right\rangle -2\left\langle \varphi_{\alpha
}|\mathbf{r}_{\alpha}|\varphi_{\alpha}\right\rangle \cdot\left\langle
\varphi_{\alpha}|\mathbf{r}_{\alpha}|\varphi_{\beta}\right\rangle \right\}
\tau_{\beta\alpha}-\left\{  \left\langle \varphi_{\alpha}|r_{\beta}%
^{2}|\varphi_{\beta}\right\rangle -2\left\langle \varphi_{\beta}%
|\mathbf{r}_{\beta}|\varphi_{\beta}\right\rangle \cdot\left\langle
\varphi_{\alpha}|\mathbf{r}_{\beta}|\varphi_{\beta}\right\rangle \right\}
\tau_{\beta\alpha}\right] \nonumber\\
&  =\sum_{\alpha\beta}\left[  \left\langle \varphi_{\alpha}|r_{\alpha}%
^{2}|\varphi_{\beta}\right\rangle -\left\langle \varphi_{\alpha}|r_{\beta}%
^{2}|\varphi_{\beta}\right\rangle \right]  \tau_{\beta\alpha}+2\left[
\left\langle \varphi_{\beta}|\mathbf{r}_{\beta}|\varphi_{\beta}\right\rangle
\cdot\left\langle \varphi_{\alpha}|\mathbf{r}_{\beta}|\varphi_{\beta
}\right\rangle -\left\langle \varphi_{\alpha}|\mathbf{r}_{\alpha}%
|\varphi_{\alpha}\right\rangle \cdot\left\langle \varphi_{\alpha}%
|\mathbf{r}_{\alpha}|\varphi_{\beta}\right\rangle \right]  \tau_{\beta\alpha
}\;.
\end{align}
As in the case of maximising \emph{U} we define an anti Hermitian matrix%

\begin{equation}
F_{\alpha\beta}=\left\langle \varphi_{\alpha}|r_{\alpha}^{2}|\varphi_{\beta
}\right\rangle -\left\langle \varphi_{\alpha}|r_{\beta}^{2}|\varphi_{\beta
}\right\rangle +2\left[  \left\langle \varphi_{\beta}|\mathbf{r}_{\beta
}|\varphi_{\beta}\right\rangle \cdot\left\langle \varphi_{\alpha}%
|\mathbf{r}_{\beta}|\varphi_{\beta}\right\rangle -\left\langle \varphi
_{\alpha}|\mathbf{r}_{\alpha}|\varphi_{\alpha}\right\rangle \cdot\left\langle
\varphi_{\alpha}|\mathbf{r}_{\alpha}|\varphi_{\beta}\right\rangle \right]  \;,
\end{equation}
and choose for the steepest descent method%

\begin{equation}
\tau=-\varepsilon F^{+}\;.
\end{equation}
The rest of the procedure is identical to the case of maximizing \emph{U}. We
can write $\mathbf{r}_{\alpha}=\mathbf{R}_{\beta}-\mathbf{R}_{\alpha
}+\mathbf{r}_{\beta}$ implying that $\left\langle \varphi_{\alpha}%
|\mathbf{r}_{\alpha}|\varphi_{\beta}\right\rangle =\left\langle \varphi
_{\alpha}|\mathbf{r}_{\beta}|\varphi_{\beta}\right\rangle $. For a one-band
case it follows that the third and fourth terms in $F_{\alpha\beta}$ vanish.
The applicability of this scheme depends crucially on the feasibility of
computing the quantities $\left\langle \varphi_{\alpha}|r_{\alpha}^{2}%
|\varphi_{\beta}\right\rangle $ and $\left\langle \varphi_{\alpha}%
|\mathbf{r}_{\alpha}|\varphi_{\beta}\right\rangle $. Once these quantities are
available the minimization process is relatively simple.

Following Ref.\cite{marzari97} the spread functional can be split according to%

\begin{equation}
\Omega=\Omega_{I}+\tilde{\Omega}\;,
\end{equation}

\begin{equation}
\Omega_{I}=\sum_{\alpha}\left[  \left\langle r^{2}\right\rangle _{\alpha}%
-\sum_{\beta}|\left\langle \varphi_{\beta}|\mathbf{r}_{\alpha}\mathbf{|}%
\varphi_{\alpha}\right\rangle |^{2}\right]  \;,
\end{equation}

\begin{equation}
\tilde{\Omega}=\sum_{\alpha}\sum_{\beta\neq\alpha}|\left\langle \varphi
_{\beta}|\mathbf{r}_{\alpha}\mathbf{|}\varphi_{\alpha}\right\rangle |^{2}\;.
\end{equation}
The quantity $\Omega_{I}$ is independent of the unitary transformation. We
could equally apply the minimization procedure to $\tilde{\Omega}$, instead of
$\Omega$.

\section{Concluding Remarks}

We have shown the usefulness of the maximally localized Wannier function as a
basis for the downfolding procedure. It is found that in transition metals the
values of the screened Coulomb interaction are in reasonable agreement with
those from previous calculations based on the LMTO-ASA. The somewhat smaller
values of the present calculations may be attributed to the more extended
nature of the Wannier orbitals compared with the more localized truncated
partial waves used in taking the matrix elements of the screened interaction
in the previous calculations. Unexpectedly we have found that for the cases we
have considered the maximally localized Wannier functions are remarkably close
to the Wannier function that maximizes the on-site Coulomb interaction.
Although we have no proof, it is quite likely that this property persists in
many other systems. This makes the maximally localized Wannier orbitals a very
suitable basis for constructing low-energy model Hamiltonians.

We have also proposed a real-space approach for constructing maximally
localized Wannier orbitals, which may be another practical procedure other
than the k-space approach. The applicability of this scheme, however, remains
to be seen.

There have been many attempts for combining first-principles methods with
many-body techniques (DMFT and its extensions, path integral renormalization
group method \cite{imada00} etc.). The present technique would be useful for
the application of these methods to real materials. In particular we have in
mind the recently developed GW+DMFT scheme to which the present maximally
localized Wannier orbitals are now being applied.

\acknowledgements We thank T. Kotani, M. Imada, N. Nakamura, R. Arita and S.
Biermann for fruitful discussions. This work was supported by Grant-in-Aid for
Scientific Research from MEXT Japan (19019013 and 19051016). We also
acknowledge computer facility from the supercomputer center at ISSP,
University of Tokyo.\newline



\begin{references}
\bibitem{gunnarsson89}
O. Gunnarsson, O.K. Andersen, O. Jepsen and J. Zaanen,
Phys. Rev. B{\bf 39}, 1708 (1989).

\bibitem{gunnarsson90}
O. Gunnarsson,
Phys. Rev. B{\bf 41}, 514 (1990).

\bibitem{aryasetiawan04}
F. Aryasetiawan, M. Imada, A. Georges, G. Kotliar, S. Biermann and A.I. Lichtenstein,
Phys. Rev. B{\bf 70}, 195104 (2004).

\bibitem{solovyev06}
I.V. Solovyev,
Phys. Rev. B{\bf 73}, 155117 (2006).

\bibitem{marzari97}
N. Marzari and D. Vanderbilt,
Phys. Rev. B\textbf{56}, 12847 (1997).

\bibitem{souza01}
I. Souza, N. Marzari and D. Vanderbilt,
Phys. Rev. B\textbf{65}, 035109 (2001).

\bibitem{andersen00}
O.K. Andersen and T. Saha-Dasgupta,
Phys. Rev. B{\bf 62}, R16219 (2000).

\bibitem{lechermann06}
F. Lechermann, A. Georges, A. Poteryaev, S. Biermann, M. Posternak, A. Yamasaki and O.K. Andersen,
Phys. Rev. B{\bf 74}, 125120 (2006).

\bibitem{kingsmith93}
R.D. King-Smith and D. Vanderbilt,
Phys. Rev. B{\bf 47}, 1651 (915).

\bibitem{giustino07}
F. Giustino, J.R. Yates, I. Souza, M.L. Cohen and S.G. Louie,
Phys. Rev. Lett. {\bf 98}, 047005 (2007).

\bibitem{wang06}
X. Wang, J.R. Yates, I. Souza and D. Vanderbilt 
Phys. Rev. B {\bf 74}, 195118 (2006).

\bibitem{yates07}
J.R. Yates, X. Wang, D. Vanderbilt and I. Souza 
Phys. Rev. B {\bf 75}, 195121 (2007).

\bibitem{miyake07}
T. Miyake, P. Zhang, M.L. Cohen and S.G. Louie,
(unpublished).

\bibitem{edmiston63}
C.E. Edmiston and K. Ruedenberg,
Rev. Mod. Phys. \textbf{35}, 457 (1963).

\bibitem{boys66}
S.F. Boys,
in \emph{Quantum Theory of Atoms, Molecules, and the Solid State}, edited by P.-O. L\"{o}wdin (Academic Press, New York, 1966).

\bibitem{anisimov91}
V.I. Anisimov, J. Zaanen, and O.K. Andersen,
Phys. Rev. B{\bf 44}, 943 (1991).

\bibitem{nakamura06}
K. Nakamura, R. Arita, Y. Yoshimoto and S. Tsuneyuki,
Phys. Rev. B{\bf 74}, 235113 (2006).

\bibitem{aryasetiawan06}
F. Aryasetiawan, K. Karlsson, O. Jepsen and U. Schonberger,
Phys. Rev. B{\bf 74}, 125106 (2006).

\bibitem {springer98}M. Springer and F. Aryasetiawan, Phys. Rev. B \textbf{57},
4364 (1998).

\bibitem {kotani00}T. Kotani, J. Phys: Condens. Matter \textbf{12, }2413 (2000).

\bibitem{solovyev05}
I.V. Solovyev and M. Imada,
Phys. Rev. B{\bf 71}, 045103 (2005).

\bibitem{cococcioni05}
M. Cococcioni and S. de Gironcoli,
Phys. Rev. B{\bf 71}, 035105 (2005).

\bibitem{kohn65}
W. Kohn and L. J. Sham,
Phys. Rev. A{\bf 140}, 1133 (1965).

\bibitem{hohenberg64}
P. Hohenberg and W. Kohn,
Phys. Rev. B{\bf 136}, 864 (1964).

\bibitem {anisimov97a}
V.I. Anisimov, F. Aryasetiawan, and A.I. Lichtenstein, 
J. Phys.: Condens. Matter \textbf{9}, 767 (1997).

\bibitem {anisimov01}For reviews, see \emph{Strong Coulomb correlations in
electronic structure calculations}, edited by V. I. Anisimov, Advances in
Condensed Material Science (Gordon and Breach, New York, 2001).

\bibitem{georges92}
A. Georges and G. Kotliar
Phys. Rev. B{\bf 45}, 6479 (1992).

\bibitem {georges96}A. Georges, G. Kotliar, W. Krauth, and M. J. Rosenberg, Rev.
Mod. Phys. \textbf{68}, 13 (1996).

\bibitem{anisimov97b}
V.I. Anisimov {\it et al.}, J. Phys.: Condens. Matter {\bf 9}, 7359 (1997).

\bibitem {sun02}P. Sun and G. Kotliar, Phys. Rev. B \textbf{66}, 085120 (2002).

\bibitem {biermann03}
S. Biermann, F. Aryasetiawan, and A. Georges,
Phys. Rev. Lett. \textbf{90}, 086402 (2003).

\bibitem{paul06}
I. Paul and G. Kotliar,
Eur. Phys. J. B{\bf 51}, 189 (2006).

\bibitem{andersen75}
O.K. Andersen,
Phys. Rev. B{\bf 12}, 3060 (1975).

\bibitem{miyake00}
T. Miyake and F. Aryasetiawan,
Phys. Rev. B{\bf 61}, 7172 (2000).

\bibitem{aryasetiawan94}
F. Aryasetiawan and O. Gunnarsson,
Phys. Rev. B{\bf 49}, 16214 (1994).

\bibitem{aryasetiawan00}
F. Aryasetiawan,
\emph{Strong Coulomb Correlations in Electronic Structure Calculations}
eds. V. I. Anisimov (Gordon and Breach, Singapore, 2000).

\bibitem{kotani02}
T. Kotani and M. van Schilfgaarde,
Solid State Commun. {\bf 121}, 461 (2002).

\bibitem{schilfgaarde06}
M. van Schilfgaarde, T. Kotani and S.V. Faleev,
Phys. Rev. B{\bf  74}, 245125 (2006).

\bibitem {imada00}M. Imada and T. Kashima, J. Phys. Soc. Jpn. \textbf{69}, 2723
(2000); T. Kashima and M. Imada, J.Phys. Soc. Jpn. \textbf{70}, 2287 (2001).
\end{references}
\end{document}